# Continuously Doping $Bi_2Sr_2CaCu_2O_{8+\delta}$ into Electron-Doped Superconductor by $CaH_2$ Annealing Method

Jin Zhao[1,2†], Yu-Lin Gan[1,2†], Guang Yang[1,2†], Yi-Gui Zhong[1,2,3], Cen-Yao Tang[1,2], Fa-Zhi Yang[1,2], Giao Ngoc Phan[1], Qiang-Tao Sui[1,2], Zhong Liu[1], Gang Li[1], Xiang-Gang Qiu[1], Qing-Hua Zhang[1], Jie Shen[1], Tian Qian[1,4], Li Lu[1], Lei Yan[1], Gen-Da Gu[5] and Hong Ding[1,2,4*]

*1 Beijing National Laboratory for Condensed Matter Physics and Institute of Physics, Chinese Academy of Sciences, Beijing 100190, China*
*2 School of Physics, University of Chinese Academy of Sciences, Beijing 100190, China*
*3 Institute for Solid State Physics, University of Tokyo, Chiba 277-8581, Japan*
*4 Songshan Lake Materials Laboratory, Dongguan 523808, China*
*5 Condensed Matter Physics and Materials Science Department, Brookhaven National Laboratory, Upton, New York 11973, USA*

*As a typical hole-doped cuprate superconductor, $Bi_2Sr_2CaCu_2O_{8+\delta}$ (Bi2212) carrier doping is mostly determined by its oxygen content. Traditional doping methods can regulate its doping level within the range of hole doping. Here we report the first application of $CaH_2$ annealing method in regulating the doping level of Bi2212. By continuously controlling the anneal time, a series of differently doped samples can be obtained. The combined experimental results of x-ray diffraction, scanning transmission electron microscopy, resistance and Hall measurements demonstrate that the $CaH_2$ induced topochemical reaction can effectively change the oxygen content of Bi2212 within a very wide range, even switching from hole doping to electron doping. We also found evidence of a low-$T_c$ superconducting phase in the electron doping side.*

The parent compounds of high-$T_c$ cuprate superconductors are usually antiferromagnetic Mott insulators with the half-filling of electron occupancy. With extra electron or hole carriers, these cuprates can become superconductors with the characteristic two-dome phase diagram. However, it has been noticed from the early days of the high-$T_c$ era that the two-dome phase diagram has a strong asymmetry between electron and hole doping. [1-6] The mystery for this asymmetry, which remains unsolved to this day, is whether it is intrinsically related to fundamental differences between electron and hole doping (i.e. due to different residing orbitals [7] or different correlation strengths caused by electron or hole charges [8]) or the structural difference between electron-doped cuprates ($T'$ structure) and hole-doped cuprates ($T$ structure). [9,10] To solve this mystery, it is desirable to have a cuprate material which can be doped by both electrons and holes, thus removing the influence of structure difference. Unfortunately, only very few systems can be doped by either holes or electrons. [11-16] Previously we made efforts along this direction by developing the ozone-vacuum annealing method. [17] While this method can *in situ* dope the surface layers of hole-doped $Bi_2Sr_2CaCu_2O_{8+\delta}$ (Bi2212) and electron-doped $La_{2-x}Ce_xCuO_4$ over a wide doping range, it fails to reach the opposite doping side. [18,19] Here we focus on the Bi2212 system again, but by using a new $CaH_2$ annealing method to further reduce the valence of Cu, which is inspired by the discovery of nickelate superconductors. [20]

In this work, we report a systematic study of regulating the doping level of Bi2212 by the $CaH_2$ annealing method (Fig. 1(a)). The original samples are optimally hole-doped Bi2212 single crystals grown by the floating-zone technique. [21] Firstly, we select high quality optimally doped

Bi2212 samples and wrap them loosely by a small piece of aluminum foil. Secondly, we take an appropriate amount of CaH$_2$ powder (0.18 gram are used in all treated samples) and put it into a quartz tube. Thirdly, we put the selected samples into the quartz tube without contacting CaH$_2$ powder and pump the tube to vacuum. Fourthly, we seal the tube after its vacuum is better than $1\times10^{-4}$ Torr, and heat it with a tube furnace. For all the samples shown in this paper unless mentioned otherwise, we adapt the heating rate of 4.5°C/min until it reaches 280°C and keep this temperature for a variable period of time (anneal time) to control the annealing sequence. Finally, we reduce the temperature to the room temperature at the same 4.5°C/min rate and then move the treated samples out of the tube. We select and cleave the samples with clean and flat surface for further experiments.

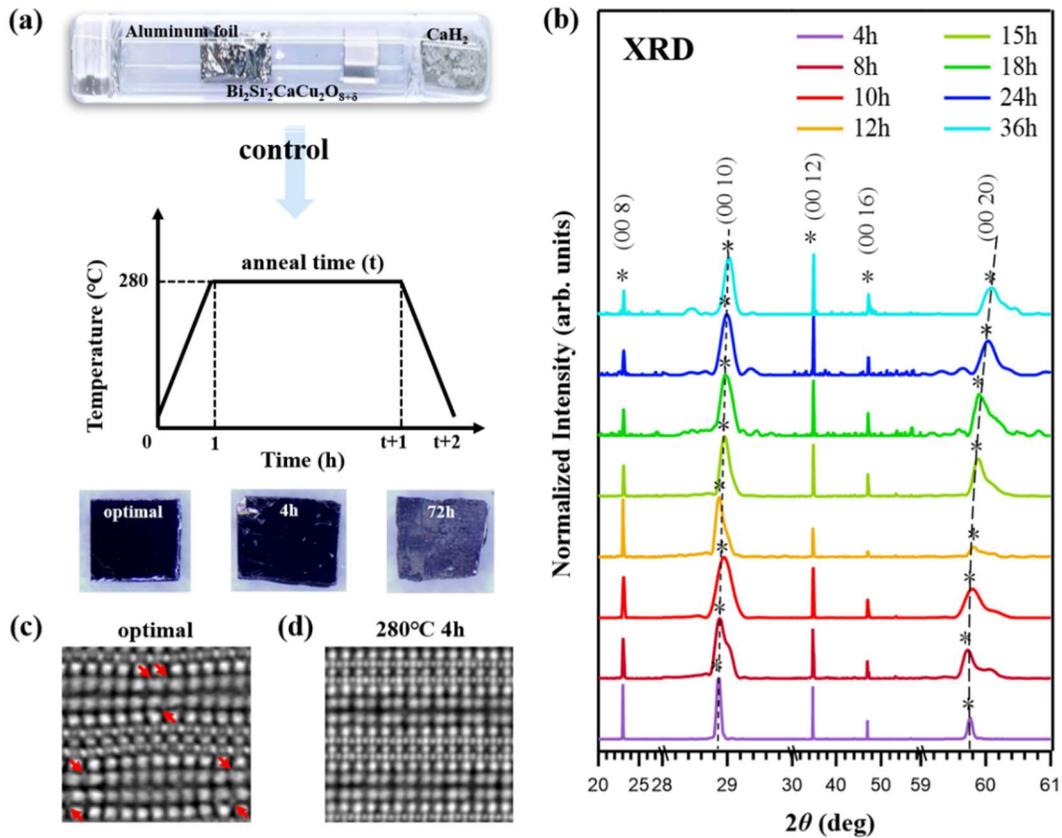

**Fig. 1.** Schematic diagram of CaH$_2$ annealing method and structural characterization of Bi2212 single crystals with different annealing time. (a) Schematic diagram of regulating the doping level of Bi2212 by the CaH$_2$ annealing method and the optical pictures of optimally doped, 4-hour annealed, and 72-hour annealed Bi2212, respectively. (b) Room-temperature XRD spectra of Bi2212 with different annealing time. All samples are annealed at 280°C, but with different annealing time. (c) Atomically resolved iDPC-STEM image of an optimally doped Bi2212 sample. (d) Same as (c) but for a Bi2212 sample annealed for 4 hours at 280°C.

With the increase of annealing time, the luster of the sample surface disappears gradually, as shown in Fig. 1(a). In order to verify that this CaH$_2$ annealing method can indeed regulate the oxygen content of Bi2212 sample, we characterized the samples by x-ray diffraction (XRD) and scanning transmission electron microscopy (STEM). Figure 1(b) shows the XRD spectra of Bi2212 with different annealing time. Within a large range of annealing time (4 hours to 36 hours), five main diffraction peaks [22] of (008), (00 10), (00 12), (00 16), (00 20) persist, indicating that the crystal structure is largely retained. With the increase of annealing time, the (00 10) diffraction peak and the (00 20) peak tend to shift to higher diffraction angles, indicating that the annealing effect reduces the *c*-axis length. The broadening of these two peaks may indicate the non-uniform reduction of the *c*-axis. Figure 1(c) shows the atomically resolved integrated differential phase contrast STEM (iDPC-STEM) image of an optimally doped Bi2212 sample. The interstitial oxygen atoms marked with red arrows are located between the BiO layers and SrO layers, similar to previous reports. [23] But after annealed for 4 hours at 280°C, the interstitial oxygen atoms disappear, as shown in Fig. 1(d). It is widely believed that the interstitial oxygen corresponds to the value of $\delta$ of the oxygen doping level, [24,25] so this result confirms that the $CaH_2$ annealing method can indeed regulate the oxygen doping level of Bi2212. In addition, the atomic arrangement of the samples after $CaH_2$ annealing is flatter than that of the optimally doped samples, as clearly seen in Figs. 1(c, d). This means that the incommensurate modulation in Bi2212 gets weaker after $CaH_2$ annealing, which would lead to a decrease of the superlattice intensity. [26]

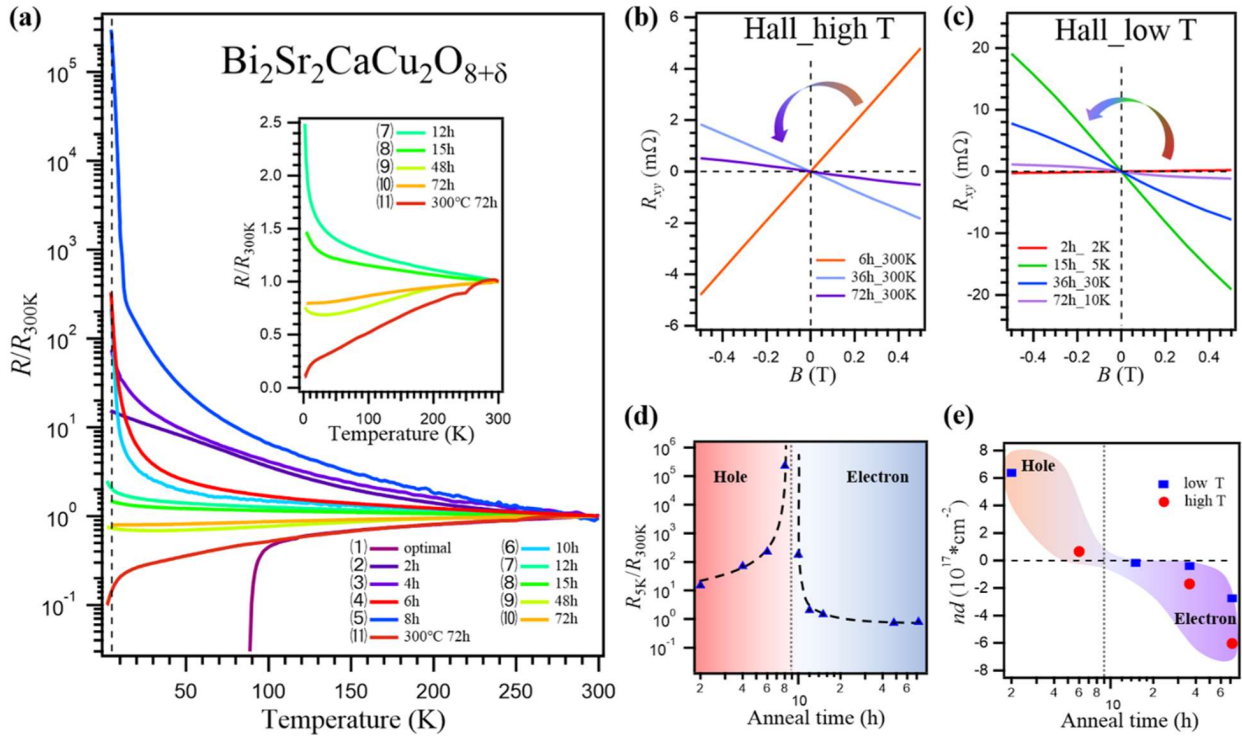

**Fig. 2.** Transport results of Bi2212 samples with different annealing time. (a) Resistance curves normalized at 300 K for Bi2212 samples with different annealing time. (b) Hall resistance of the samples annealed for different time, which is measured by Van der Pauw method at the room temperature (300K). (c) Hall resistance measured at low temperatures (≤30K). (d) Resistance ratio of 5K and 300K as a function of the annealing time. (e) Product of Hall carrier concentration (*n*) and sample thickness (*d*) as a function of the annealing time. To better visualization, the annealing time (2 hours to 72 hours) is commuted to the logarithm form.

To determine to what extent the doping level of Bi2212 system can be regulated by this $CaH_2$ annealing method, we perform transport measurements, as shown in Fig. 2. We measure the resistances of samples by the four-wire method [27] and normalize them with their corresponding resistances at 300K (Fig. 2(a)). The resistance of the samples starts to exhibit the insulator-like behavior when annealed for 2 hours, and reaches the maximum around an 8-hour annealing time, then decreases with further increase of annealing time. This behavior suggests that the samples change from the metallic/superconducting phase to the insulating phase and revert to another metallic phase. Furthermore, we measure the Hall resistance of the samples with different annealing time by the Van der Pauw method. [28] Figures 2(b) and 2(c) show the Hall resistance measured at the room temperature (300K) and at the low temperatures (≤30K), respectively. With the increase of annealing time, the sign of Hall resistance gradually changes from positive to negative, indicating that the main carrier type changes from hole to electron. To better quantify the change of resistance, we show the resistance ratios of 5K and 300K of the samples with different annealing time in Fig. 2(d). Apparently, the increase in resistance is because that the samples approach an insulating parent phase as the hole doping level decreases. With further reduction of oxygen content, the resistance of the samples gradually decreases, which indicates that the Bi2212 system may enter the electron-doped region. Considering the thicknesses ($d$) of all the samples are approximately the same, Figure 2(e) actually shows the variation trend of Hall carrier concentration ($n$) with annealing time, which is calculated from Figs. 2(b, c). These transport results indicate that it is possible to change the doping type from hole to electron in the Bi2212 system by the CaH$_2$ annealing method.

Next, it is desirable to find out whether the metallic state in the electron-doped side can become superconducting at a lower temperature. Therefore, we preform the lower temperature transport measurements by using a dilution refrigerator which can reach a lowest temperature of 10mK. Remarkably, we find that a new superconducting phase appeared in some samples with a long annealing time, as shown in Fig. 3(a), which displays three samples annealed for more than 77 hours at 290°C with a much slower heating and cooling rate of 0.45°C/min. The onset temperatures of the superconductivity ($T_{c,\,onset}$) are ~1.23K, 1.4K and 1.44K, respectively, and the temperatures with zero resistance ($T_{c,0}$) are ~0.86K, 0.94K and 1.1K, respectively. The small deviation between $T_{c,\,onset}$ and $T_{c,0}$ indicates the sharp superconducting transitions. The detailed superconducting properties of this new superconductor will be reported in a separate paper. We also measure the Hall resistance of the sample annealed for 84 hours at 290°C (Fig. 3(b)), and find that its carriers are indeed mainly electrons. Further transport experiments shown in the supplementary material (Fig. S1) strongly suggest that the superconductivity is not derived from residual hole-doped Bi2212. Furthermore, we examine the values of $T_c$ of the elements (Bi, Sr, Ca, Cu, O) contained in the Bi2212 sample and some multi-component compounds of these elements. [29-34] $T_c$ of Bi bulk under the ambient pressure is below 0.53mK while $T_c$ of cylindrical Bi nanowire is below 1.3K. [35] $T_c$ of Bi element can reach 8.2K under pressure up to 8.1 GPa while $T_c$ of its single oxide $\delta$-Bi$_2$O$_3$ is 5.8K.[36,37] As for the multi-component compounds of these elements, there are also some combinations can become high-$T_c$ superconducting materials, such as Bi$_2$Sr$_2$CuO$_{6+\delta}$ (Bi2201) and Bi$_2$Sr$_2$Ca$_2$Cu$_3$O$_{10+\delta}$ (Bi2223). We also found trace of a low-$T_c$ superconducting phase when the Bi2212 system is doped into the electron doping side. Considering the growth technology, pressure environment, transport properties, XRD patterns, etc., the superconductors listed above are not east to form. Nevertheless, we cannot completely exclude the possibility of trace amount. Therefore, this electron-doped superconductivity may arise from the electron-doped Bi2212 phase, or from the secondary impurity phase.

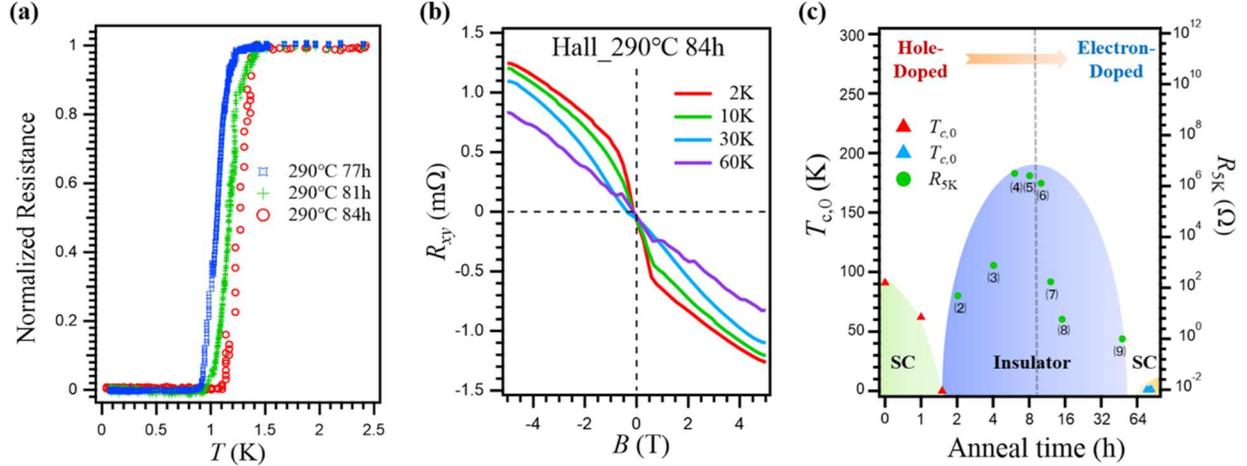

**Fig. 3.** Ultra-low-temperature transport results of the heavily annealed superconductors. (a) Resistance curves normalized at 2.5 K with three different annealing time of 77, 81, and 84 hours, respectively. (b) Hall resistance of the sample annealed for 84 hours at 290°C. (c) A phase diagram of the Bi2212 system reached by the $CaH_2$ annealing method. The triangle symbols refer the $T_{c,0}$ and the green circles are the corresponding resistance values measured at 5K in Fig. 2(a).

We summarize our results in Fig. 3(c), which roughly sketches a phase diagram that the Bi2212 samples reached by this new $CaH_2$ annealing process. The ability of reaching the insulting phase or even the electron side in the Bi2212 system may help us to solve the mystery of electron-hole asymmetry in the cuprates. In fact, we have conducted some preliminary ARPES results over this new phase diagram, which will be presented in a separate paper. However, one caveat for this method needs to be mentioned here: during the anneal process, hydrogen coming out of $CaH_2$ would combine with oxygen extracted out of Bi2212 and forms water, and the water vapor will likely degrade the Bi2212 single crystal as it comes out of the sample. To improve the single crystal quality, one can try to optimize the hydrogenation conditions or use thin film samples. The reason why we could not get higher $T_c$ in the electron-doped samples might be that the heavily annealed samples tend to have poorer quality. In fact, the reason why electron doping is so difficult to achieve in Bi2212 and most hole-doped systems is likely due to the fact that their charge transfer gaps are larger than the electron-doped systems such as $Nd_{2-x}Ce_xCuO_4$, so a brute force method such as the $CaH_2$ annealing process is needed to reach the electron-doping side.

In conclusion, we develop the $CaH_2$ annealing method to regulate the doped carrier of Bi2212 single crystal samples over an unprecedently wide doping range, and may even reach the superconducting electron-doped phase. This will likely open a new venue for systematically studying the cuprate systems from both electron and hole sides, from which new results would provide new insights on some of the most fundamental issues the high-$T_c$ cuprate superconductors are facing, including the electron-hole asymmetry, the pseudogap, and the high-$T_c$ superconducting mechanism.

*Acknowledgments*. We thank Xing-Chen Guo, Zhe Zheng, Feng Ran and Ren-Jie Zhang for technical support, and Zheng-Yu Weng, Fu-Chun Zhang and Tao Xiang for helpful discussions. This work was supported by the National Natural Science Foundation of China (No.11888101, U1832202), the Chinese Academy of Sciences (QYZDB-SSW-SLH043, XDB33000000), the K. C. Wong Education Foundation (GJTD-2018-01), and the Informatization Plan of Chinese Academy of Science (CAS-WX2021SF-0102). This work was also supported by the Synergetic Extreme Condition User Facility (SECUF). Y.L.G. was supported by China Postdoctoral Science

Foundation (2020M680726, YJ20200325). G.D.G. was supported by US DOE (DE-SC0010526, DE-SC0012704).

† These authors contributed equally to this work.
* Corresponding author. Email: dingh@iphy.ac.cn
PACS: 74.72.-h;74.25.F-;74.72.Ek;74.25.Dw

# Supplementary Material for "Continuously Doping $Bi_2Sr_2CaCu_2O_{8+\delta}$ into Electron-Doped Superconductor by $CaH_2$ Annealing Method"

Jin Zhao[1,2,†], Yu-Lin Gan[1,2,†], Guang Yang[1,2,†], Yi-Gui Zhong[1,2,3], Cen-Yao Tang[1,2], Fa-Zhi Yang[1,2], Giao Ngoc Phan[1], Qiang-Tao Sui[1,2], Zhong Liu[1], Gang Li[1], Xiang-Gang Qiu[1], Qing-Hua Zhang[1], Jie Shen[1], Tian Qian[1,4], Li Lu[1], Lei Yan[1], Gen-Da Gu[5] and Hong Ding[1,2,4*]

*1 Beijing National Laboratory for Condensed Matter Physics and Institute of Physics, Chinese Academy of Sciences, Beijing 100190, China*
*2 School of Physics, University of Chinese Academy of Sciences, Beijing 100190, China*
*3 Institute for Solid State Physics, University of Tokyo, Chiba 277-8581, Japan*
*4 Songshan Lake Materials Laboratory, Dongguan 523808, China*
*5 Condensed Matter Physics and Materials Science Department, Brookhaven National Laboratory, Upton, New York 11973, USA*

**The supplementary materials include the following contents:**

**I.  Oxidation reaction of Bi2212 samples with different anneal time**

**II. Transport experiments of electron-doped samples with the highest $T_c$ achieved**

**III. XRD patterns of some superconducting materials**

**I.  Oxidation reaction of Bi2212 samples with different anneal time**

When we anneal the Bi2212 sample for 84 hours at 290°C, we find a new superconducting phase. To verify whether this superconductivity is a remnant of hole-doped superconductivity, we perform the oxidation reaction experiment for some samples (Fig. S1). We select some samples with different anneal time, such as the optimally doped samples, the samples annealed for 2 hours and 84 hours by CaH$_2$ annealing method. $T_{c,0}$ of the optimally doped sample is about 91K, while $T_{c,0}$ of the samples annealed for 84 hours is about 1.1K, and the sample annealed for 2 hours is known as an insulator at the hole-doped side. In the oxidation reaction process, high-purity oxygen is added into the tube furnace, where the selected samples were placed, at a flow rate of 2 L/min and heated at 550°C for 3 hours with a heating and cooling rate of 5°C/min. After the oxidation reaction, the original optimally doped sample enters the over-doped superconducting region ($T_{c,0}$ ~78K), and the sample annealed for 2 hours by the CaH$_2$ annealing method returns to the underdoped superconducting region ($T_{c,0}$ ~82K), but the sample annealed for 84 hours becomes an insulator. This oxidation experiment provides evidence that the new superconducting phase is not a remnant of hole-doped superconductivity, otherwise it would remain superconducting rather than insulating after the oxidation reaction.

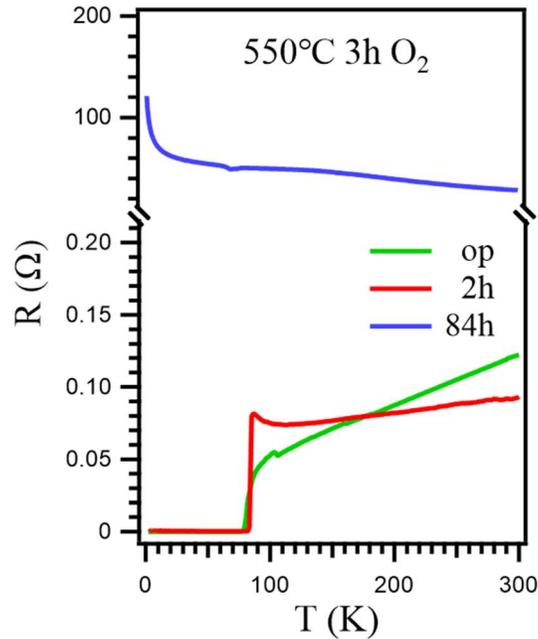

**Fig. S1** Transport results of different samples after oxidation for 3 hours at 550°C. Samples before adding oxygen correspond to different hydrogenation annealing times. The green, red and blue line corresponds to the optimally doped sample, the sample annealed for 2 hours, and the sample annealed for 84 hours by the CaH$_2$ annealing method, respectively.

## II. Transport experiments of electron-doped samples with the highest T$_c$ achieved

The values of $T_c$ for the new superconducting phase are below 1.5K, which may be due to poorer quality of the single crystal, but we once achieved the superconductivity with $T_{c,0}$ of 6K, and its Hall results also suggest the electron doping nature (Fig. S2)

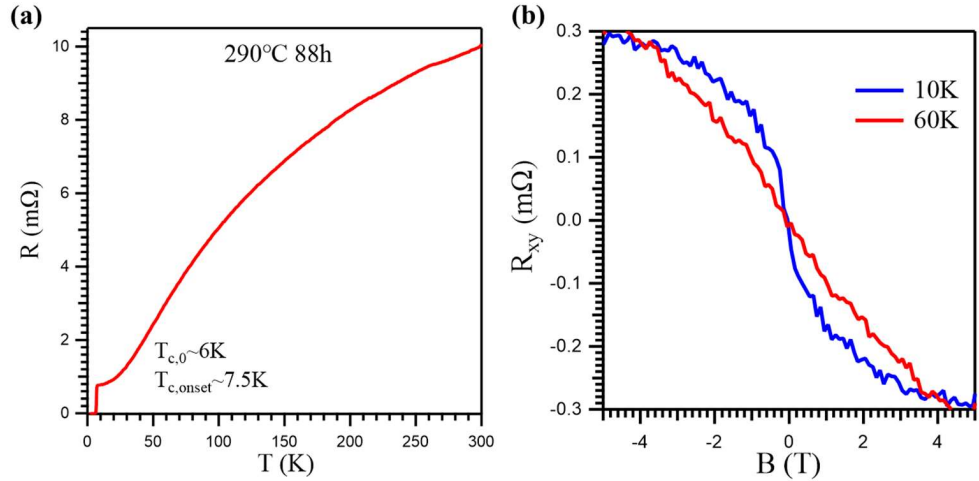

**Fig. S2** Transport results of a sample annealed for 88 hours at 290°C. (a) Resistance curve (b) Hall resistance of this sample.

### III. XRD patterns of some superconducting materials

We measure the XRD pattern of the sample annealed for 84 hours at 290°C, and find that its peak positions shift slightly to higher angles and the half-width also increases. This means that although the quality of the single crystal has degraded, the structure of Bi2212 is still basically maintained. The peaks of Bi single element (PDF#441246) are not visible in the XRD, which means that the new superconducting phase does not likely originate from Bi nanowires, although we cannot exclude the possibility of trace amount.

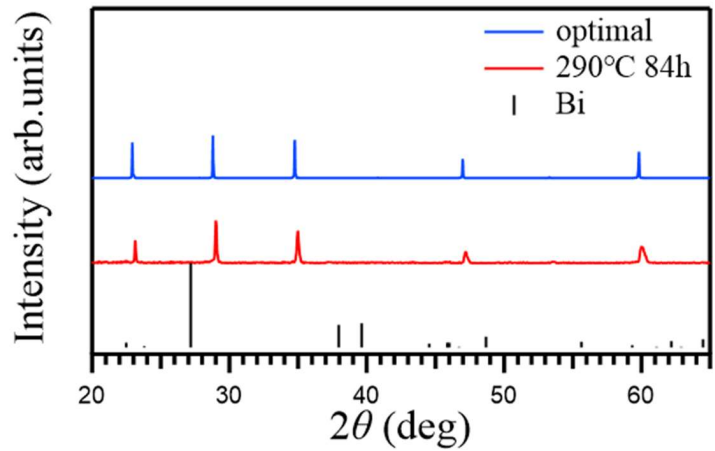

**Fig. S3** XRD patterns of the optimally doped sample (blue line) and the sample annealed for 84 hours at 290°C (red line). The black line is from the XRD patterns of Bismuth (PDF card#441246).